\documentclass[english, preprint,showpacs,preprintnumbers,amsmath,amssymb]{revtex4}
\usepackage[T1]{fontenc}
\usepackage[latin1]{inputenc}
\usepackage{amssymb}

\makeatletter

\usepackage{babel}
\makeatother
\begin{document}

\title{PT-symmetric Models with $O(N)$ Symmetry}

\author{Peter N. Meisinger and Michael C. Ogilvie}

\affiliation{Department of Physics, Washington University, St. Louis, MO 63130
USA}

\begin{abstract}
We construct $PT$-symmetric quantum mechanical models with an
$O(N)$-symmetric interaction
term of the form $-g\left(\vec{x}^{2}\right)^{2}/N$. Using functional
integral methods, we find the equivalent Hermitian model, which has
several unusual features. The effective potential obtained in the
large-$N$ limit of the Hermitian form is shown to be identical to
the form obtained from the original $PT$-symmetric model using familiar
constraint field methods. The analogous constraint field prescription
in four dimensions suggests that $-g\left(\vec{\phi}^{2}\right)^{2}/N$ 
$PT$-symmetric scalar field theories
are asymptotically free.
\end{abstract}


\maketitle

\section{Introduction}

Since the initial discovery of $PT$ symmetry \cite{Bender:1998ke},
 there has been
considerable progress in expanding both the number of $PT$-symmetric
models and our knowledge of their properties \cite{Bender:2005tb,Bender:2007nj}.
Although the construction
of $PT$-symmetric matrix models has proved to be relatively straightforward
\cite{Meisinger:2007jx}, 
the construction of $PT$-symmetric models with fields transforming
as vectors under $O(N)$ appears to have more difficult technical
issues. Nevertheless, the development of scalar field theory models
with vector symmetry is crucial to the possible relevance of $PT$-symmetry
in particle physics. On the other hand, there has been substantial
progress recently in understanding the relation of the
one-component $-\lambda x^{4}$
 model to its equivalent Hermitian form 
\cite{Bender:2006wt,Jones:2006et}, as well as 
work on the relation of $O(N)$-symmetric Hermitian
models to one-component $PT$-symmetric models
\cite{Andrianov:2007vt}.
In this paper, we extend this work to
a $PT$-symmetric model with $O(N)$ symmetry, deriving the Hermitian form for all
values of $N$. We construct the large-$N$ limit, and show that
this limit can be obtained using the familiar method of constraint 
fields, but in a much less rigorous way.

Consider a model with Euclidean Lagrangian given by\begin{equation}
L_{E}=\sum_{j=1}^{N}\left[\frac{1}{2}\left(\partial_{t}x_{j}\right)^{2}+\frac{1}{2}m^{2}x_{j}^{2}-\lambda x_{j}^{4}\right]-\frac{g}{N}\left(\sum_{j=1}^{N}x_{j}^{2}\right)^{2}\end{equation}
where $g$ and $\lambda$ are non-negative. When $g=0$, we have $N$
decoupled one-dimensional systems; for $\lambda=0$, we have a model
with $O(N)$ symmetry. When both $g$ and $\lambda$ are non-zero,
the model has only an $S_{N}$ permutation symmetry. From the standpoint
of $PT$ symmetry, the interaction terms can be considered as members
of a family of $PT$-invariant interactions\begin{equation}
-\lambda\sum_{j=1}^{N}\left(-ix_{j}\right)^{2p}-\frac{g}{N}\left(-\sum_{j=1}^{N}x_{j}^{2}\right)^{q}\end{equation}
which are invariant under $PT$ symmetry. This class of models
is well-defined for $p=q=1$, and must be defined for $p,q>1$ by
an appropriate analytic continuation of the $x_{j}$ as necessary
 \cite{Bender:1998ke}.

It is convenient to consider this model as a subset of a larger class
of models, with a Lagrangian of the form\begin{equation}
L_{E}=\sum_{j=1}^{N}\left[\frac{1}{2}\left(\partial_{t}x_{j}\right)^{2}+\frac{1}{2}m^{2}x_{j}^{2}\right]-\sum_{j,k=1}^{N}x_{j}^{2}\Lambda_{jk}x_{k}^{2}\end{equation}
The classical stability of the potential for large $x_{j}$ is governed
by the eigenvalues of $\Lambda$. For the model of particular interest
to us, \begin{equation}
\Lambda=\lambda I+gP\end{equation}
where $P$ is the projector\begin{equation}
P=\frac{1}{N}\left(\begin{array}{ccc}
1 & 1 & 1\\
1 & 1 & ..\\
1 & .. & ..\end{array}\right)\end{equation}
satisfying $P^{2}=P$. The decomposition $\Lambda=\lambda\left(I-P\right)+\left(g+\lambda\right)P$
shows that $\Lambda$ has one eigenvalue $g+\lambda$ and $N-1$ eigenvalues
with value $\lambda$. The eigenvalue $g+\lambda$ is associated with
variations in $\vec{x}^{2},$ i.e., variations in the radial direction.

\section{Equivalence to Hermitian models}

We will analyze the case where all eigenvalues of $\Lambda$ are positive
using functional integration. With the substitution
\begin{equation}
x_{j}\rightarrow-2i\sqrt{c_{j}+i\psi_{j}}\end{equation}
familiar from the one-component case, $L_{E}$ becomes \begin{equation}
L_{E}=\sum_{j}\left[\frac{1}{2}\frac{\left(\partial_{t}\psi_{j}\right)^{2}}{(c_{j}+i\psi_{j})}-2m^{2}(c_{j}+i\psi_{j})\right]-16\sum_{jk}\Lambda_{jk}(c_{j}+i\psi_{j})(c_{k}+i\psi_{k}).\end{equation}
The generating function for the model is given by \begin{equation}
Z=\int\prod_{j}\frac{[d\psi_{j}]}{\sqrt{\det(c_{j}+i\psi_{j})}}\exp\left[-\int dt\left[L_{E}-\sum_{j}\frac{1}{32}\left(\frac{1}{c_{j}+i\psi_{j}}\right)\right]\right]\end{equation}
where the change of variables has generated both a functional determinant
and additional term, formally of order $\hbar^{2}$, in the action.
As pointed out in \cite{Jones:2006et}, both terms are required to obtain correct results
in the functional integral formalism.

The functional determinant may be written as 
\begin{equation}
\prod_{j}\frac{1}{\det\left[\sqrt{c_{j}+i\psi_{j}}\right]}=\int\prod_{j}\left[dh_{j}\right]\exp\left\{ -\int dt\,\left[\frac{1}{2}\left(c_{j}+i\psi_{j}\right)\left(h_{j}-\frac{\dot{i\psi_{j}}+1/4}{c_{j}+i\psi_{j}}\right)^{2}\right]\right\} \end{equation}
which introduces a new set of fields $h_{j}$. The derivation proceeds
as in the single-variable case. After integration by parts on the
$h_{j}\dot{\psi_{j}}$ terms, and adding and subtracting total derivatives,
the functional integral over the $\psi_{j}$ fields can be carried
out exactly. The integral is both local and quadratic, and requires
that the matrix $\Lambda$ have positive eigenvalues for convergence.
The result of this integration is \begin{equation}
Z=\int\prod_{n}\left[dh_{n}\right]\exp\left\{- \int dt\,\left[\frac{1}{64}\sum_{jk}\Lambda_{jk}^{-1}\left[\frac{1}{2}h_{j}^{2}+\dot{h}_{j}-2m^{2}\right]\left[\frac{1}{2}h_{k}^{2}+\dot{h}_{k}-2m^{2}\right]-\sum_{j}\frac{1}{4}h_{j}\right]\right\}. \end{equation}
After discarding total derivatives, we obtain\begin{equation}
Z=\int\prod_{n}\left[dh_{n}\right]\exp\left[-\int dt\,\frac{1}{64}\sum_{jk}\Lambda_{jk}^{-1}\left[\dot{h}_{j}\dot{h}_{k}+\frac{1}{4}\left(h_{j}^{2}-4m^{2}\right)\left(h_{k}^{2}-4m^{2}\right)\right]-\sum_{j}\frac{1}{4}h_{j}\right]\end{equation}
 which gives the Hermitian form for our general $PT$-symmetric model
with $N$ fields.

In the particular case we are interested in, we have \begin{equation}
\Lambda^{-1}=\frac{1}{\lambda}\left(I-P\right)+\frac{1}{g+\lambda}P.\end{equation}
The Lagrangian may be written as\begin{equation}
L_{E}=\frac{1}{64\lambda}\sum_{j}\left[\dot{h}_{j}^{2}+\frac{1}{4}\left(h_{j}^{2}-4m^{2}\right)^{2}\right]-\frac{g}{64N\lambda\left(g+\lambda\right)}\left[\left(\sum_{j}\dot{h}_{j}\right)^{2}+\frac{1}{4}\left(\sum_{j}\left(h_{j}^{2}-4m^{2}\right)\right)^{2}\right]-\frac{1}{4}\sum_{j}h_{j}.\end{equation}
 It is helpful to immediately rescale all the fields as $h_{j}\rightarrow\sqrt{32\lambda}h_{j}$:\begin{equation}
L_{E}=\sum_{j}\left[\frac{1}{2}\dot{h}_{j}^{2}+4\lambda\left(h_{j}^{2}-\frac{m^{2}}{8\lambda}\right)^{2}\right]-\frac{g}{N\left(g+\lambda\right)}\left[\frac{1}{2}\left(\sum_{j}\dot{h}_{j}\right)^{2}+4\lambda\left(\sum_{j}\left(h_{j}^{2}-\frac{m^{2}}{8\lambda}\right)\right)^{2}\right]-\sqrt{2\lambda}\sum_{j}h_{j}.\end{equation}
 At this point, the $S_{N}$ permutation symmetry is still manifest,
and it clear that the field $\sum_{j}h_{j}$ plays a special role.

In order to understand the strategy for rewriting the model in a form
in which the limit $\lambda\rightarrow0$ can easily be taken, it
is useful to work out explicitly the case of $N=2$ first. It is apparent
that a rotation of the fields will be desirable. We define suggestively
new fields $\sigma$ and $\pi$ given by\begin{equation}
\begin{array}{c}
h_{1}=\frac{1}{\sqrt{2}}\left(\sigma+\pi\right)\\
h_{2}=\frac{1}{\sqrt{2}}\left(\sigma-\pi\right)\end{array}.\end{equation}
 After some algebra and the rescaling\begin{equation}
\sigma\rightarrow\sqrt{\frac{g+\lambda}{\lambda}}\sigma\end{equation}
we arrive at\begin{equation}
L_{E}=\frac{1}{2}\dot{\sigma}^{2}+\frac{1}{2}\dot{\pi}^{2}-m^{2}\sigma^{2}-\frac{\lambda m^{2}}{g+\lambda}\pi^{2}+2\left(g+\lambda\right)\sigma^{4}+\frac{2\lambda^{2}}{g+\lambda}\pi^{4}+\left(8g+12\lambda\right)\sigma^{2}\pi^{2}-2\sqrt{g+\lambda}\sigma.\end{equation}
Notice the natural hierarchy between the masses for $\lambda\ll g$.
The $O(2)$ symmetric limit of the original $PT$-symmetric model
is obtained in the limit $\lambda\rightarrow0$, where we have\begin{equation}
L_{E}=\frac{1}{2}\dot{\sigma}^{2}+\frac{1}{2}\dot{\pi}^{2}-m^{2}\sigma^{2}+2g\sigma^{4}+8g\sigma^{2}\pi^{2}-2\sqrt{g}\sigma.\end{equation}
The field $\pi$ has no mass term, indicating its relation to the
angular degrees of freedom in the original Lagrangian. However, radiative
corrections generate a mass for the $\pi$ field via the the $\sigma^{2}\pi^{2}$
interaction. As in the one-component case, there is a linear anomaly term, but
only for $\sigma$.

We now turn to the more difficult case of the $\lambda\rightarrow0$
limit for arbitrary $N$. As before, we introduce a field $\sigma$
defined by\begin{equation}
\sigma=\frac{1}{\sqrt{N}}\sum_{j}h_{j}\end{equation}
as well as a set of $N-1$ fields $\pi_{k}$ with $k=1,..,N-1$ related
to the $h_{j}$ fields by a rotation so that $\sigma^{2}+\vec{\pi}^{2}=\vec{h}^{2}$.
Each field $h_{j}$ can be written as\begin{equation}
h_{j}=\frac{1}{\sqrt{N}}\sigma+\tilde{h}_{j}\end{equation}
 where $\sum_{j}\tilde{h}_{j}=0$. This property is crucial in eliminating
a term in $L_{E}$ which diverges as $\lambda^{-1/2}$ as $\lambda\rightarrow0$. 
The Lagrangian now can be written as\begin{equation}
L_{E}=\frac{1}{2}\dot{\sigma}^{2}+\sum_{j}\frac{1}{2}\dot{\pi}_{j}^{2}-m^{2}\left(\sigma^{2}+\vec{\pi}^{2}\right)+4\lambda\sum_{j}h_{j}^{4}+\frac{4}{N}\left(\frac{\lambda^{2}}{g+\lambda}-\lambda\right)\left(\sigma^{2}+\vec{\pi}^{2}-\frac{Nm^{2}}{8\lambda}\right)^{2}-\sqrt{2\lambda N}\sigma.\end{equation}
The rescaling $\sigma\rightarrow\sqrt{\left(g+\lambda\right)/\lambda}\sigma$
plus some careful algebra yields the $\lambda\rightarrow0$ limit as\begin{equation}
L_{E}=\frac{1}{2}\dot{\sigma}^{2}+\frac{1}{2}\dot{\vec{\pi}}^{2}-m^{2}\sigma^{2}+\frac{4g}{N}\sigma^{4}+\frac{16g}{N}\sigma^{2}\vec{\pi}^{2}-\sqrt{2gN}\sigma\end{equation}
 which agrees with our previous result for $N=2$ , and agrees with
the known result for a single degree of freedom if we take $N=1$
and drop the $\vec{\pi}$ field altogether. This is a Hermitian form
of the $PT$-symmetric anharmonic oscillator with $O(N)$ symmetry,
derived as the limit of a $PT$-symmetric model with $S_{N}$ symmetry.
The Hermitian form has several novel features. Note that both the
$S_{N}$ and $O(N)$ symmetries are no longer manifest, but there
is an explicit $O(N-1)$ symmetry associated with rotations of the
$\vec{\pi}$ field. As in the $N=2$ case, there is no mass term for
the $\vec{\pi}$ field. Furthermore, there is no $\left(\vec{\pi}^{2}\right)^{2}$
term, although there is a $\vec{\pi}^{2}\sigma^{2}$ interaction.
The anomaly term again involves only $\sigma$, and breaks the symmetry
$\sigma\rightarrow-\sigma$ possessed by the rest of the Lagrangian.
Analyzing the Lagrangian at the classical level, we see that if $m^{2}>0$
, the $\sigma$ field is moving in a double-well potential, perturbed
by the anomaly so that $\left\langle \sigma\right\rangle >0$.$ $
On the other hand, if $m^{2}<0$, $\sigma$ moves in a single-well
anharmonic oscillator, again with the linear anomaly term making $\left\langle \sigma\right\rangle >0$.
In either case, the $\vec{\pi}^{2}\sigma^{2}$ interaction will generate
a mass for the $\vec{\pi}$ field. All of this is consistent with
the association of $\sigma$ and $\vec{\pi}$ with the radial and
angular degrees of freedom, respectively, in the original $PT$-symmetric
model.

\section{Large-$N$ Limit}

We will defer a more detailed discussion of this model for finite
$N$, and turn to its large-$N$ limit. One more rescaling $\sigma\rightarrow\sqrt{N}\sigma$
gives the Lagrangian\begin{equation}
L_{E}=\frac{N}{2}\dot{\sigma}^{2}+\frac{1}{2}\dot{\vec{\pi}}^{2}-Nm^{2}\sigma^{2}+4gN\sigma^{4}+16g\sigma^{2}\vec{\pi}^{2}-N\sqrt{2g}\sigma.\end{equation}
 We see that the anomaly term survives in the large-$N$ limit, unlike
the matrix model case \cite{Meisinger:2007jx}. After integrating over the $N-1$ 
components of the $\vec{\pi}$
field, we have the large-$N$ effective potential $V_{eff}$ for
$\sigma$:\begin{equation}
V_{eff}/N=-m^{2}\sigma^{2}+4g\sigma^{4}+\frac{1}{2}\sqrt{32g\sigma^{2}}-\sqrt{2g}\sigma.\end{equation}
It is striking that the anomaly term has virtually the same form as
the zero-point energy of the $\vec{\pi}$ field. The anomaly term
breaks the discrete $\sigma\rightarrow-\sigma$ symmetry of the other
terms of the Lagrangian, and always favors $\sigma\geq0$. The effective
potential has a global minimum with $\sigma$ positive for $m^{2}>3\,2^{1/3}g^{2/3}$.
For $m^{2}<3\,2^{1/3}g^{2/3}$, there does not appear to be a stable
solution with $\sigma>0$, and $\sigma=0$ is the stable solution
to leading order in the $1/N$ expansion. This change in the behavior
of the effective potential as $m^{2}$ is varied is not seen in the
corresponding Hermitian model \cite{Coleman:1974jh}, and indicates
a need for care in analyzing the model. Based on our preliminary analysis
of the Hermitian form for finite $N$, we believe that this behavior
is associated with the large-$N$ limit, and does not indicate a fundamental
restriction on $m^{2}$.

The large-$N$ effective potential was derived from a Lagrangian with
unusual properties, associated with the Hermitian form of the original
model. It is therefore surprising that, once the form of the large-$N$
effective potential is known, it can be derived heuristically in a
more conventional way. We start from the $O(N)$-symmetric Lagrangian\begin{equation}
L_{E}=\sum_{j=1}^{N}\left[\frac{1}{2}\left(\partial_{t}x_{j}\right)^{2}+\frac{1}{2}m^{2}x_{j}^{2}\right]-\frac{g}{N}\left(\sum_{j=1}^{N}x_{j}^{2}\right)^{2}\end{equation}
 and add a quadratic term in a constraint field $\rho$\begin{equation}
L_{E}\rightarrow L_{E}+\frac{g}{N}\left(\frac{2N\rho}{g}+\sum_{j=1}^{N}x_{j}^{2}-\frac{Nm^{2}}{4g}\right)^{2}\end{equation}
yielding\begin{equation}
L_{E}=\sum_{j=1}^{N}\left[\frac{1}{2}\left(\partial_{t}x_{j}\right)^{2}+4\rho x_{j}^{2}\right]+\frac{4N\rho^{2}}{g}-\frac{Nm^{2}\rho}{g}+\frac{Nm^{4}}{16g}.\end{equation}
If we integrate over $x_{j}$ in a completely conventional way, we
obtain the large-$N$ effective potential\begin{equation}
V_{eff}/N=\frac{4\rho^{2}}{g}-\frac{m^{2}\rho}{g}+\sqrt{2\rho}+\frac{m^{4}}{16g}.\end{equation}
This is essentially identical to our previous expression after identifying
$\rho=g\sigma^{2}$. However, we lack a fundmental justification for
this approach. We know that great care must be taken in specifying
the contour of integration in typical $PT$-symmetric models, yet
the $x_{j}$ fields were integrated over quite conventionally. If
this approach has validity, it seems likely that the choice of integration
contours for $\rho$ and $\vec{x}$ is crucial. However, only the
saddle point matters to leading order in $1/N$, so it is possible
for this heuristic derivation to be correct even though we lack a
direct, complete treatment of the original $PT$-symmetric model.

\section{$PT$-symmetric field theory}

If we boldly apply the constraint field approach to a $PT$-symmetric
field theory with a $-g\left(\vec{\phi}^{2}\right)^{2}$ interaction
in $d$ dimensions, we obtain the effective potential\begin{equation}
V_{eff}/N=\frac{4\rho^{2}}{g}-\frac{m^{2}\rho}{g}+\frac{m^{4}}{16g}+\frac{1}{2}\int\frac{d^{d}k}{\left(2\pi\right)^{d}}\ln\left[k^{2}+8\rho\right].\end{equation}
Models of this type were rejected decades ago \cite{Coleman:1974jh}
because of stability
concerns at both the classical and quantum levels,
although there were early indications that
such theories were in fact sensible \cite{Andrianov:1981wu}.
Within the framework
of $PT$-symmetric models, such stability issues cannot be addressed
without understanding the boundary conditions on the functional integrals.
However, it is straightforward to check that renormalization of $g$
in $d=4$ gives an asymptotically free theory, with beta function
$\beta=-g^{2}/2\pi^{2}$ in the large-$N$ limit. If $PT$-symmetric
scalar field theories exist in four dimensions and 
are indeed asymptotically free, the possible
implications for particle physics are large, and provide ample justification
for further work.

Finally, we note that the construction we have used for $PT$-symmetric
models with $O(N)$ symmetry is not the only one possible. For example,
we can consider our original model with $g>0$ but $\lambda<0$, so
that only the $O(N)$ symmetric term is unconventional. It would be
interesting to know if the $\lambda\rightarrow0$ limit of this class
of models can be used to define a $PT$-symmetric model with $O(N)$
symmetry, and if so, if it is equivalent to the one studied here.

\bibliography{PT_Vector_MS}

\end{document}